\documentclass[12pt]{article}
\usepackage{amsfonts}
\usepackage{amssymb}
\usepackage{epsfig}
\newcommand{\beq}{\begin{equation}}
\newcommand{\eeq}{\end{equation}}
\renewcommand{\Re}{\mathop{\mathrm{Re}}}
\renewcommand{\Im}{\mathop{\mathrm{Im}}}
\newcommand{\sgn}{\mathop{\mathrm{sgn}}}
\newcommand{\tr}{\mathop{\mathrm{tr}}}
\newcounter{one}
\newtheorem{theorem}{Theorem}
\newtheorem{lemma}{Lemma}
\newtheorem{proposition}{Proposition}

\begin{document}
\begin{flushright}
\parbox{9cm}{
\begin{center}
This paper is dedicated
to S.P.Novikov, \\ 
on the occasion of his 65th birthday.
\end{center}
}
\end{flushright}

\begin{center}
{\Large\bf
The initial boundary value problem on the segment
for the Nonlinear Schr\"odinger equation;
the algebro-geometric approach. \Roman{one}}
\end{center}
\begin{center}
P.G.Grinevich\footnote{The main part of this work was fulfilled during the
author's visit to the University of Roma ``La Sapienza'' in 2002, supported by
a fellowship of the Italian Ministry of Foreign Affairs, organized by the
Landau Network-Centro Volta. The author was also supported by the RFBR
grant No. 01-01-0874A.}

L.D.Landau Institute for Theoretical Physics,

Kosygina 2, 117940, Russia,

e-mail: pgg@landau.ac.ru

\medskip

P.M.Santini\footnote{The author acknowledges partial support from the INFN,
Sezione di Roma ``La Sapienza'' and from the research funds of the cultural agreement
between the Landau Institute of Moscow and the University of Roma
``La Sapienza''.}

Dipartimento di Fisica, Universit\`a di Roma ``La Sapienza'' \&

INFN, Sezione di Roma 1,

Piazzale A.Moro 2, 00185, Roma, Italia.

e-mail: paolo.santini@roma1.infn.it

\end{center}

\abstract{ This is the first of a series of papers devoted to the study of
classical {\bf initial-boundary value problems} of Dirichlet, Neumann and mixed
type for the Nonlinear Schr\"odinger equation on the segment. Considering proper 
periodic discontinuous extensions of the profile, generated by suitable point-like 
sources, we show that the above boundary value  problems can be rewritten as nonlinear
dynamical systems for suitable sets of algebro-geometric spectral data, generalizing 
the classical Dubrovin equations. 

In this paper we consider, as a first illustration of the above method, the case of the 
Dirichlet problem on the segment with zero-boundary value at one end, and we show that 
the corresponding dynamical system for the spectral data can be written as a system of ODEs 
with algebraic right-hand side.}

\section{Introduction}

In the natural phenomena, evolutionary processes very often take place
in a domain bounded in space and time. At the boundary the system has
a non-trivial interaction with the external world or experimental
environment. The mathematical description of such situations is given
by {\bf initial boundary value} (IBV) problems for evolution equations.
Since soliton equations are largely applicable, IBV problems for them
are of large interest from both mathematical and physical points of
view. It is the case, for instance, of the IBV problems for the
Nonlinear Schr\"odinger (NLS) equation
\beq
\label{NLS}
iq_t=q_{xx}\pm 2q^2\bar q, \ \ \ q=q(x,t), \ \ \ q\in{\mathbb C}
\eeq
which describes the complex amplitude modulation of a
quasi-monochromatic wave packet in strongly dispersive and weakly
nonlinear media, and then arises in many natural contexts, like in
fiber optics, plasma physics and fluid dynamics.

Two IVB problems for soliton  equations are well-studied and
understood.

1) The initial problem on the line with rapidly decreasing boundary
conditions at infinity (the Cauchy problem on the line). This problem
is solved using the scattering transform, first introduced for
the Korteweg-de Vries (KdV) equation in \cite{GGKM} and later used
to solve the NLS equation in \cite{ZS71}. In this case the scattering
data evolve linearly and can be treated as a nonlinear analog of
the continuous Fourier data.

2) The IBV problem on the segment [0,L] with periodic boundary
conditions. Its solution is based on the so-called finite-gap approach,
first introduced in \cite{Nov1}.

More general IBV problems (for example with  Dirichlet, Neumann,
mixed and Robin boundary conditions) for soliton equations are
believed to be non-integrable. Only for special choices of the boundary
values the problems remain integrable \cite{AS},\cite{Skl} and the
above two standard integration procedures with appropriate reductions on the 
spectral data can be applied ( see \cite{AS} for the semi-line case and 
\cite{ItsB} for the segment one; see also \cite{Taras} for the semi-line case 
using the approach of \cite{Skl}); but these cases 
cover only a limited number of interesting physical applications.

Applying the scattering transform to more general IBV
problems on the semi-line or on the segment, one  meets the
following basic difficulty: the time evolution of the scattering
matrix depends not only on prescribed boundary values, but also
on unknown ones. Several strategies
have been developed to overcome such difficulty (a list of them
can be found at the end of this introduction). One of the possible
strategies, motivated by the classical solutions of the Dirichlet and
Neumann problems for second order linear PDEs by the sine and cosine 
Fourier transforms, consists in transforming the given IBV problem
into a certain Cauchy problem on the whole line by introducing a proper
extension of the profile such that the known boundary values are encoded into
external point sources and the unknown ones disappear from
this new formulation. Using this strategy, the Dirichlet problem
on the semi-line  for the NLS equation was investigated in \cite{Fok2}.

More recently, the same strategy was applied to the Dirichlet and Neumann
problems on the semi-line for the NLS equation in \cite{DMS1} using
a different formalism. In \cite{DMS1} the time evolution of the
scattering matrix $\tilde S(\lambda,t)$ of the transformed problem
on the whole line is expressed in terms of the scattering matrix
$S(\lambda,t)$ of the original semi-line problem and of the
given boundary values. To close the system, one uses the non-local
connection $\tilde S(\lambda,t)= S^{-1}(-\lambda,t) \sigma S(\lambda,t)$ ,
where $\sigma=1$ for the Dirichlet case, and $\sigma=\sigma_3$ for the
Neumann case ($\sigma_3$ is the Pauli matrix). Due to the
analyticity properties of the scattering matrix $S(\lambda,t)$,
the above non-local relation can be resolved for $S(\lambda,t)$
using the appropriate Riemann-Hilbert problem.

Motivated by this result, we have recently begun a
research activity with  the purpose of applying the same
methodology to the IBV problems on the segment [0,L] for the
NLS equation. In particular, we have found that:

a) the Dirichlet problem is conveniently
reformulated as an odd $2L$-periodic Cauchy problem on
the line with $\delta'$ sources located at the points $nL$ with
$n\in\mathbb Z$.

b) The Neumann problem is reformulated as an even $2L$-periodic
Cauchy problem on the line with $\delta$ sources located at the
points $nL$.

c) The mixed problem with, for instance, $u(0,t)$ and
$u_x(L,t)$ given, is reformulated as an odd $4L$-periodic Cauchy
problem such that $q(2L-x,t)=q(x,t), \ 0\le x \le L$, with $\delta'$
sources located at the points $2nL$ and $\delta$ sources at the
points $(2n+1)L$. Here and later we denote by $u(x,t)$ the NLS field
in the segment $[0,L]$ and by $q(x,t)$  its extension to the whole line.

\begin{center}
\mbox{\epsfxsize=10cm \epsffile{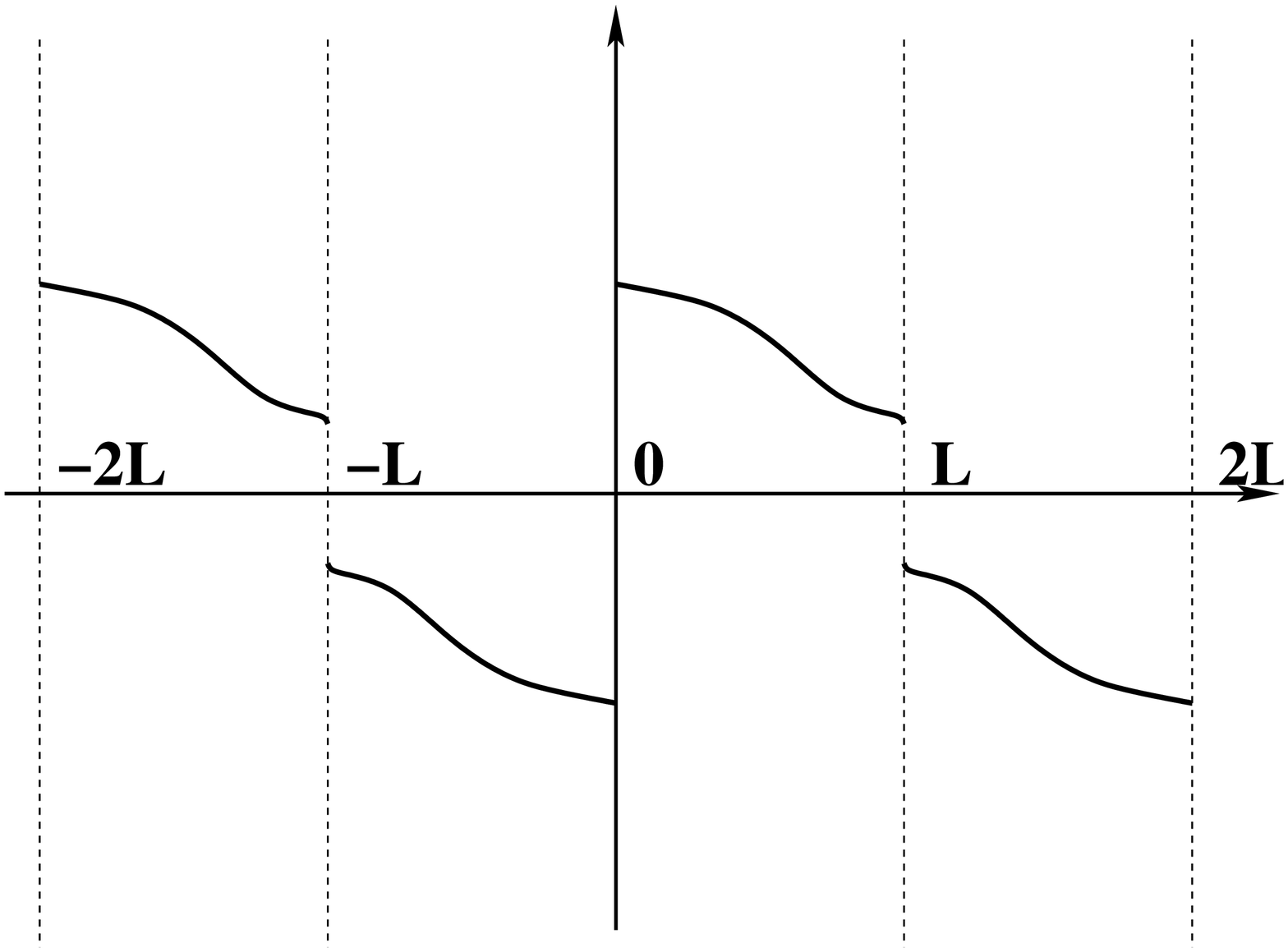}}

Fig 1.

Odd $2L$-periodic extension for the Dirichlet IBV problem.

\end{center}

Through these new formulations of the original IBV problems, we have been able to
obtain the following results.

1) We have eliminated from the formalism all the unknown boundary values.

2) We have transformed the original IBV problem into a Cauchy
problem for periodic profiles, for which the powerful tools of
the finite-gap method are at hand.

3) We have used the algebro-geometric tools of the finite gap method to
reduce the original IBV problem to a nonlinear
system of ordinary differential equations describing the time evolution of
the spectral data of the transformed periodic problem with forcings.
This system gives a satisfactory analytic description of the
IBV problem under investigation.

Since the IBV problem on the segment is equivalent to a periodic problem
for non-smooth profiles produced by point-like sources, then
the corresponding spectral theory presents the following novel
features with respect to the smooth case.

1) The lengths of the spectral gaps decay slowly
in the high-energy limit (they are of order $1/\lambda$ for the
Dirichlet problem and of order $1/\lambda^2$ for the Neumann one),
while, in the smooth case, the gaps lengths decay faster than
any power of $\lambda$.

2) The number of gaps is always infinite and the finite-gap approximation
converges rather slowly.

3) The branch points are not conserved any more
and their time evolution must be included into the set Dubrovin
equations.

4) The evolution of the spectral curve is described in terms
of the isospectral deformations suggested in \cite{GS}.

In this paper we illustrate the above method on the
following particular Dirichlet problem for the defocussing NLS equation:
\beq
\label{DfNLS}
iu_t=u_{xx}-2u^2\bar u, \ \ \ u=u(x,t), \ \ \ u\in{\mathbb C},
\ \ \ 0\le x\le L, \ \ \ t \ge 0.
\eeq
\beq
\label{IBV1}  
u(0,t)=v_0(t), \ \ u(L,t)=v_L(t)=0, \ \ u(x,0)=u_0(x),\ \ \ 0\le x\le L,
\eeq
where $v_0(t)$ and $u_0(x)$ satisfy the obvious matching conditions
$v(0)=u_0(0)$, $u_0(L)=0$. The choice of the defocussing case is due to
its simpler analytic structure (the auxiliary spectral problem is
self-adjoint) and the choice of 0-boundary at $x=L$ simplifies
the Dubrovin equations. From the point of view of the method used,
these simplifications are not essential and
the solution of the general Dirichlet, Neumann and mixed  IBV problems,
for both the focussing and defocussing NLS, will be presented in a more 
detailed forthcoming paper.

As it was previously indicated, to study the IBV problem (\ref{DfNLS}),
(\ref{IBV1}) it is convenient to introduce the odd $2L$-periodic
extension of $u(x,t)$:
\beq
\label{extension}
\begin{array}{c}
q(x,t)=\sum\limits_{n\in {\mathbb Z}}u(x-2nL,t) H(x-2nL)H((2n+1)L-x)-
\\
u(2nL-x,) H(2nL-x)H(x-(2n-1)L),\ \ x\in{\mathbb R}, \ \ t>0.
\end{array}
\eeq
This extension satisfies the following odd
$2L$-periodic Cauchy problem on the line with $\delta'$ forcings
\beq
\label{DfNLS2}
iq_t=q_{xx}-2q^2\bar q - 2 v_0(t) \sum\limits_{n\in {\mathbb Z}}
\delta'(x-2nL),\ \ x\in{\mathbb R}, \ \ t>0,
\eeq
\beq
\label{IBV2}
\begin{array}{c}
q(x,0)=\sum\limits_{n\in {\mathbb Z}} u_0(x-2nL) H(x-2nL)H((2n+1)L-x)-
\\
u_0(2nL-x) H(2nL-x)H(x-(2n-1)L),
\end{array}
\eeq
which will be the subject of investigation of the following sections.
We show, in particular, that unlike the case of generic forcings, 
which lead to a highly non-local formalism, the point-like sources arising 
from the IBV problem (\ref{DfNLS}), (\ref{IBV1}) lead to a system of ODEs for
the spectral data which can be written in local form after a proper extension 
of the phase space. This picture is consistent with the mild non-locality found 
in the semi-line case \cite{DMS1}.

The paper is organized as follows. In \S 2 we summarize the main results concerning
the finite gap theory of the defocussing NLS equation for smooth periodic profiles.
In \S 3 we present the spectral characterization of the discontinuous periodic
profiles arising from the Dirichlet problem. In \S4 we derive the
nonlinear system of ordinary differential equations describing the time evolution
of the spectral data.

We end this introductory section with a list of
other approaches to the study of IBV problems for soliton equations,
developed during the last few years.
In \cite{Sabat} an ``elbow scattering'' in the $(x,t)$- plane is presented
to deal with the semi-line problem for KdV, leading to
Gel'fand - Levitan - Marchenko formulations. In \cite{Fokas1,Fokas2} a
different approach, based on a simultaneous x-t spectral transform,
has been introduced and rigorously developed in \cite{FIS,Fokas5},
to solve IBV problems for soliton equations on the semi-line. 
We were recently informed \cite{FI2} that this method was also 
extended to the segment case. This method allows for a 
rigorous asymptotics \cite{FI} and captures in a natural way the known 
cases of linearizable IBV problems. In \cite{DMS2} and 
\cite{DMS1} two alternative approaches to the study of IBV problems for soliton equations on the
segment and on the semi-line have been presented. In the first method, applicable to both
the semi-line and segment cases, the unknown boundary values are expressed in terms of elements 
of the scattering matrix $S(\lambda ,t)$, 
thus obtaining a nonlinear integro-differential evolution equation for $S$.
In the second method one applies a suitable linear operator to $S_tS^{-1}$ to
eliminate the unknown boundary data. It is also shown that the resulting time evolution of 
$S$, nonlinear and mildly nonlocal,
coincides with that obtained applying the point source strategy of this paper \cite{DMS1}.
Also the formalism presented in \cite{DMS1} allows for rigorous asymptotics and
captures in a natural way the case of linearizable IBV problems.
Some integrable boundary conditions for soliton equations and their connection
with symmetries have been investigated in \cite{Habibullin} and in references therein quoted.
In some non-generic cases of soliton equations corresponding to singular
dispersion relations, like the stimulated
Raman scattering (SRS) equations and the sine Gordon (SG)
equation in light-cone coordinates, the evolution equation of the scattering
matrix does not contain unknown boundary data.
The SG equation on the semi-line has been treated using the
x-t spectral transform \cite{Fokas1}; the SRS and the SG equations on
the semi-line have also been treated using a more traditional x-
transform method respectively in \cite{LM} and in \cite{LS}; the x-
spectral data used in this last approach satisfy a nonlinear evolution
equation of Riccati type.
Apart from the simultaneous $x-t$ transform, all the above approaches
are based on the traditional IST \cite{libri}. A different approach, 
based on the Kac-Moody representation for the SG equation in laboratory 
coordinates, was used in \cite{Bel} to solve the Robin problem on 
the semi-line; it uses critically the finite-speed character of the 
equation.

The spectral formalism for studying forced soliton equations has been
developed by several authors, especially in connection to the
theory of perturbations (see, for instance, \cite{Newell} and \cite{Gerd})).
Soliton equations subjected to one-point-like source were investigated 
in \cite{FA}.
Soliton equations perturbed by slowly varying forcings were investigated by averaging
procedures in \cite{EFMS} and \cite{Kr2}. We finally remark that IBV problems and 
forced problems for C - integrable equations have been considered in
\cite{Delillo}, \cite{AL}.

\section{Periodic spectral transform for the defocussing NLS}
\subsection{Direct periodic spectral transform}
\label{Section2.1}
The Lax pair for the defocussing NLS equation:

\beq
\label{S2-DfNLS}
iq_t=q_{xx}-2q^2\bar q, \ \ \ q=q(x,t),
\eeq
was found in \cite{ZS71}. We use its zero-curvature form:
\beq
\label{S2-lax1x}
\frac{\partial \Psi(\lambda,x,t)}{\partial x}=
U(\lambda,x,t) \Psi(\lambda,x,t),
\eeq
\beq
\label{S2-lax1t}
\frac{\partial \Psi(\lambda,x,t)}{\partial t}=
V(\lambda,x,t) \Psi(\lambda,x,t),
\eeq
where $\Psi(\lambda,x,t)$ is a 2-component vector:
\beq
\label{S2-lax2}
\Psi(\lambda,x,t)=\left[\begin{array}{c} \psi^1(\lambda,x,t)\\
\psi^2(\lambda,x,t) \end{array} \right]
\eeq
and $U(\lambda,x,t)$,  $V(\lambda,x,t)$
are the following $2\times2$ matrices:

\beq
\label{S2-lax3x}
U(\lambda,x,t)=
\left[\begin{array}{cc}i\lambda & iq(x,t) \\ -i\bar q(x,t) & -i\lambda
\end{array}\right],
\eeq
\beq
\label{S2-lax3t}
V(\lambda,x,t)=2\lambda U(\lambda,x,t)+
\left[\begin{array}{cc}iq\bar q & q_x \\ \bar q_x & -iq\bar q  \end{array}
\right].
\eeq

Equation (\ref{S2-lax1x}) can be rewritten as the following spectral
problem
\beq
\label{S2-lax4x}
{\cal L}\Psi(\lambda,x,t)=\lambda \Psi(\lambda,x,t),
\eeq
where
\beq
\label{S2-lax5x}
{\cal L}=\left[\begin{array}{cc}-i\partial_x & -q(x,t) \\
- \bar q(x,t) & i\partial_x  \end{array}\right].
\eeq

The first symmetry of the NLS hierarchy is the infinitesimal gauge
\beq
\label{S2-NLS_0}
q_{t_0}=iq.
\eeq
Its zero-curvature representation has the following form:
\beq
\label{S2-lax0}
\frac{\partial \Psi}{\partial x}=U \Psi, \ \
\frac{\partial \Psi}{\partial t_0}=V_0 \Psi,
\eeq
where
\beq
\label{S2-lax00}
V_0= \frac{i}{2}\sigma_3= \left[\begin{array}{cc}\frac i2  & 0 \\ 0
& -\frac i2 \end{array}\right].
\eeq

Assume, that our potential is periodic with the period $2L$:
\beq
q(x+2L,t)\equiv q(x,t)
\eeq

Denote by  $\hat T(\lambda,x,a,t)$ the fundamental solution of
(\ref{S2-lax1x})
\beq
\label{S2-T1}
\frac{\partial \hat T(\lambda,x,a,t)}{\partial x}= U(\lambda,x,t)
\hat T(\lambda,x,a,t), \ \ \hat T(\lambda,a,a,t)=
\left( \begin{array}{cc} 1 & 0 \\ 0 & 1 \end{array} \right).
\eeq
We shall use:
\beq
\label{S2-T2}
\hat T(\lambda,b,a,t)=
\hat\Psi(\lambda,b,t)\hat\Psi^{-1}(\lambda,a,t)
\eeq
where $\hat\Psi(\lambda,x,t)$ is an arbitrary matrix
solution of (\ref{S2-lax1x}) with non-zero determinant.

The matrix
\beq
\hat T(\lambda,t)=\hat T(\lambda,2L,0,t)
\eeq
is called {\bf monodromy matrix}.

Let us associate spectral data with $q(x,t)$ (we treat
${\cal L}$ as an ordinary differential operator in $x$
depending on the parameter $t$).
\begin{lemma}
The following spectral problems are self-adjoint:
\begin{enumerate}
\item \label{S2-sp1} Main problem: the spectral problem for the operator
${\cal L}$ on the whole line in the space $L^2(\mathbb R)$.
\item \label{S2-sp2} Auxiliary problem: the spectral problem for the operator
${\cal L}$ on the interval $[0,2L]$ with the boundary conditions:
\beq
\psi^1(0)+\psi^2(0)=0, \ \ \psi^1(2L)+\psi^2(2L)=0.
\eeq
\end{enumerate}
\end{lemma}

The spectrum of the main problem is the union of closed intervals
of the real line. The complement of this spectrum is the union of
open intervals called gaps. Let us denote the boundary points
of the gap number $j$ by $E_{2j}$ and $E_{2j+1}$, $E_{2j}<E_{2j+1}$.
The spectrum of the auxiliary problem is discrete. Each interval
$[E_{2j},E_{2j+1}]$ (the closure of the gap $j$) contains exactly
one point of the auxiliary spectrum.

For generic potentials the gaps can be enumerated by integers
$-\infty<j<\infty$, and for sufficiently large $j$ the gap number
$j$ is located near the point $\pi j/2 L$. Each point $\lambda_k$
of the auxiliary spectrum is associated with a gap, and we can assume
that $E_{2k}\le\lambda_k\le E_{2k+1}$. The lengths of gaps tends to
$0$ as $|j|\rightarrow\infty$. The decay rate depends on the
smoothness of $q(x,t)$ and coincides with the decay rate of the Fourier
coefficients of the potential.

For non-generic potentials the spectrum of the auxiliary problem
can also be enumerated so that $\lambda_k\rightarrow \pi k/2 L$
$|k|\rightarrow\infty$, but some of these points lie inside the
spectrum of the main problem. Such potentials can be interpreted
as the results of ``closing'' some gaps: $E_{2j+1}\rightarrow E_{2j}$.
To construct the inverse transform for the self-adjoint problem
discussed in our text, only the points of the auxiliary spectrum $\lambda_j$
associated with gaps are essential.

These spectra possess the following simple characterization in terms
of the monodromy matrix $\hat T(\lambda,t)$:

\begin{lemma}
\label{S2-l-spectrum}
\begin{enumerate}
\item A real number $\lambda$ lies in the spectrum of the main
problem if and only if $|\tr\hat T(\lambda,t)|\le2$.
\item A number $\lambda_k$ is a point of the auxiliary spectrum
if and only if the vector
$$
\left(\begin{array}{c} 1 \\ -1
\end{array}\right)
$$
is an eigenvector of $\hat T(\lambda,t)$, or equivalently,
$$
T_{11}(\lambda,t)+T_{21}(\lambda,t)-T_{12}(\lambda,t)-T_{22}(\lambda,t)=0.
$$
\end{enumerate}
\end{lemma}

The Bloch eigenfunction $\Psi(\gamma,x)$ is, by definition, the common
eigenfunction of ${\cal L}$ and of the shift operator

\beq
\label{bloch1}
{\cal L }\Psi(\gamma,x,t)=\lambda \Psi(\gamma,x,t), \ \ \
\Psi(\gamma,x+2L,t)=e^{2iLp(\gamma)}\Psi(\gamma,x,t),
\eeq
where $\gamma$ is a point of a Riemann surface $\Gamma$, which is a
two-sheeted covering of the $\lambda$-plane, 
${\cal P}$ denotes the projection from $\Gamma$ to the $\lambda$-plane:
${\cal P}\gamma=\lambda$
and $e^{2iLp(\gamma)}$ is an eigenvalue of the monodromy matrix. 
The branch points of $\Gamma$ are real and coincide with the ends 
$E_{j}$ of the gaps.
Denote by $a_k$ the oval lying over the interval $[E_{2k},E_{2j+k}]$
and by $\sigma$ the transposition of sheets.

The function $p(\gamma)$ is defined modulo $(\pi/L) n$,
$n\in {\mathbb Z}$. It is called the {\bf quasimomentum}.
It is odd with respect to the transposition of sheets:
\beq
p(\sigma\gamma)\equiv -p(\gamma), \ \ \mbox{mod}(\pi/L).
\eeq

The differential of the quasimomentum $dp=\frac{dp}{d\lambda} d\lambda$
is holomorphic at the finite part of the spectral curve $\Gamma$. It has
exactly 2 zeroes in each oval $a_k$: $\alpha_k^+$ and $\alpha_k^-$,
$\sigma \alpha_k^+ = \alpha_k^- $:
\beq
\label{S2-alphas}
\left.\frac{\partial p(\gamma,t)}
{\partial \lambda}\right|_{\gamma=\alpha^{\pm}_k}=0,
\eeq
\beq
\label{S2-alphaplus}
\Im p(\alpha_k^+)>0.
\eeq
Denote by $\alpha_k$ the projection of these zeroes to the $\lambda$-plane.

Equation (\ref{bloch1}) defines $\Psi(\gamma,x,t)$ up to a constant
factor. In our text we use the following normalization:
\beq
\label{S2-bloch1.1}
\Phi(\gamma,0,0)=1,
\eeq
where
\beq
\label{S2-bloch2}
\Phi(\gamma,x,t)=\psi^1(\gamma,x,t)+\psi^2(\gamma,x,t).
\eeq
It is also convenient to introduce the following function:

\beq
\Xi(\gamma,x,t)=(\psi^1(\gamma,x,t)-\psi^2(\gamma,x,t))
\Phi^{-1}(\gamma,x,t).
\eeq
By definition:
\beq
\label{S2-defpsi}
\Psi(\gamma,x,t)=\left[\begin{array}{c}
\frac {1+\Xi(\gamma,x,t)}{2}  \\
\frac{1-\Xi(\gamma,x,t)}{2} \end{array} \right]
\Phi(\gamma,x,t).
\eeq

For each $x$ and $t$, function $\Phi(\gamma,x,t)$ has exactly one zero
$\gamma_k(x,t)$ on the oval $a_k$. The poles $\gamma_k$ of $\Phi(\gamma,x,t)$
do not depend on $x$, $t$, they lie on the ovals over the intervals
$[E_{2k},E_{2j+k}]$ and each oval contains exactly one pole $\gamma_k$,
with $\gamma_k=\gamma_k(0,0)$. Denote by
$\lambda_k$ the projections of the points $\gamma_k$ to the $\lambda$-plane.
From the definition it follows that they are the points of the auxiliary
spectrum corresponding to the gaps.

Hence the following spectral data are associated with a periodic
potential of the defocussing NLS equation.
\begin{enumerate}
\item A collection of real branch points $E_k$ (or, equivalently, the
spectral curve $\Gamma$).
\item The divisor of poles $\gamma_k$ in $\Gamma$.
\end{enumerate}
These data uniquely determine the potential $q(x,t)$.

\subsection{The periodic inverse problem. Finite-gap potentials}

A potential $q(x,t)$ is called {\bf finite-gap} if the surface $\Gamma$
is algebraic or, equivalently, if the genus $g$ of $\Gamma$ is finite.
In the finite-gap case we define $\Gamma$ by the algebraic equation:
\beq
\label{S2-alg1}
\mu^2=R(\lambda), \ \ \  R(\lambda)=\prod\limits_{k=0}^{2g+1}
(\lambda-E_k).
\eeq
By definition a point $\gamma$ of $\Gamma$ is a pair of complex numbers
$\gamma=(\lambda,\mu)$ satisfying (\ref{S2-alg1}). Therefore the poles
and the zeroes of $\Phi(\lambda,x,t)$ are pairs of complex numbers
$\gamma_k=(\lambda_k,\mu_k)$ and complex functions
$\gamma_k(x,t)=(\lambda_k(x,t),\mu_k(x,t))$ respectively, satisfying
(\ref{S2-alg1}). Here $0\le k\le g$.

The role of the finite-gap potentials in the soliton theory is
analogous to the role of finite Fourier series in the linear theory.

In the finite-gap case the function $\Xi(\gamma,x,t)$ has the following
analytic properties:
\begin{enumerate}
\item $\Xi(\gamma,x,t)$ is meromorphic in $\gamma$ on the spectral
curve $\Gamma$.
\item $\Xi(\gamma,x,t)$ has at most first-order poles at the divisor
points $\gamma_k(x,t)$ and no other singularities.
\item $\Xi(\gamma,x,t)\rightarrow\pm1$ as $\gamma\rightarrow\pm\infty$.
\end{enumerate}

\begin{lemma}
The properties formulated above completely determine the function
$\Xi(\gamma,x,t)$. The explicit formula for $\Xi(\gamma,x,t)$ is given by:
\beq
\label{S2-defXi}
\Xi(\gamma,x,t)=\frac{\mu}{\prod\limits_{k=0}^{g}(\lambda-\lambda_k(x,t))}+
\sum\limits_{k=0}^{g}\frac{r_k(x,t)}{2(\lambda-\lambda_k(x,t))},
\eeq
\end{lemma}
where
\beq
\label{S2-r_k}
r_k(x,t)=2 \frac{\mu_k(x,t)} {\prod\limits_{{j=0,\ldots,g}\atop{j\ne k} }
(\lambda_k(x,t)-\lambda_j(x,t))}.
\eeq

One of the possible procedures for reconstructing potentials by
the spectral data is the following. The ends $E_k$ of the gaps 
are constants of motion, i.e. they are $x,t$-independent.
The divisor of zeroes $\gamma_k(x,t)$ satisfy the following 
systems of first-order ordinary differential
equations in $x$ and $t$ (Dubrovin equations):
\beq
\label{S2-Dubrx}
\frac{\partial\lambda_k(x,t)}{\partial x}=
-i\left[\frac12\sum\limits_{j=0}^{2g+1}E_j -
\sum\limits_{{j=0,\ldots,g}\atop{j\ne k} } \lambda_j(x,t)
\right] r_k(x,t),
\eeq
\beq
\label{S2-Dubrt}
\frac{\partial\lambda_k(x,t)}{\partial t}=
-2i\left[\frac12\sum\limits_{j=0}^{2g+1}E_j^2 -
\sum\limits_{{j=0,\ldots,g}\atop{j\ne k} } \lambda_j^2(x,t)
\right] r_k(x,t).
\eeq
Solving equations (\ref{S2-Dubrx}), (\ref{S2-Dubrt}) for the initial
data $\lambda_k(0,0)=\lambda_k$, $\mu_k(0,0)=\mu_k$, one obtains
the divisor of zeroes $\gamma_k(x,t)$ for all $x$, $t$. Using
the reconstruction formula
\beq
\label{S2-potent}
q(x,t)=\sum\limits_{k=0}^{g}\left[\lambda_k(x,t)-\frac12(E_{2k}+E_{2k+1})
-\frac{r_k(x,t)}{2}\right]
\eeq
one gets the potential $q(x,t)$.

In the derivation of the Dubrovin equations one assumes that the spectral 
curve has a finite genus. These equations can also be used for the 
infinite-gap case, but it is necessary to be careful with the associated 
convergence problems.

We use also Dubrovin equations for the infinitesimal gauge
(\ref{S2-NLS_0}):
\beq
\label{S2-Dubrph}
\frac{\partial\lambda_k(x,t)}{\partial t_0}=
-\frac{i}{2} r_k(x,t)
\eeq

In the finite-gap case, the Dubrovin equations can be explicitly linearized
using the Abel transform and the potentials can be written in terms of
Riemann $\Theta$-functions associated with the curve $\Gamma$. But
we do not use this property in our text.

In the finite-gap case the $\lambda$-derivative of the quasimomentum
differential reads as:
\beq
\label{S2-dp}
\frac{\partial p(\gamma)}{\partial \lambda} = 
\frac{\prod\limits_{k=0}^g (\lambda-\alpha_k)}{\mu}
\eeq

{\bf Remark.} The algebro-geometrical (or finite-gap) solutions of the
KdV equation were first constructed in \cite{Nov1}, where the zero-curvature 
representation was also first introduced. The complete
finite-gap theory for the periodic Schr\"odinger operator
was developed in \cite{DN1,DN2,D1,IM1,L,McK-VM} (see the survey article
\cite{DMN} and the books \cite{libri}, \cite{BBEIM}). A pure algebraic 
formulation of the finite-gap integration procedure and its 
generalization for 2+1 dimensional systems were
obtained in \cite{Kr3}.

Finite-gap solutions of the defocusing NLS equation were first
constructed in \cite{Its}. The first $\theta$-functional formulas
for the focussing NLS were written in \cite{IK}. A detailed study of the
finite-gap NLS solutions can be found in \cite{Pre85}.
An interpretation of generic periodic 1-d Schr\"odinger potentials in
terms of $\Theta$-functions of infinitely many variables was
suggested in \cite{MT}. A generalization of this approach to
generic matrix ordinary differential operators was suggested in
\cite{Sch96}. The spectral transform for the operator (\ref{S2-lax1x})
with potentials in the Hilbert space $L^2([0,2L])$ was developed
in \cite{Korot}.

In the finite-gap case we mark a point of $\gamma\in\Gamma$ by a pair of
complex numbers $\gamma=(\lambda,\mu)$. But, to study the limit
$g\rightarrow\infty$, it is convenient to renormalize $\mu$.
Assume that we have a point inside the k-th gap:
$\lambda\in{\mathbb R}$, $E_{2k}\le\lambda\le E_{2k+1}$. Then
we define a new variable $\tilde\mu$ by
\beq
\label{S2-tmu-def}
\mu=i\tilde\mu\sqrt{\left|
\prod\limits_{{j\ne2k}\atop{j\ne2k+1}}(\lambda-E_j) \right|}
\sgn\left(\prod\limits_{j\ne k}(\lambda-\lambda_j) \right),
\eeq
\beq
\label{S2-tmu-rel}
\tilde\mu^2=(E_{2k+1}-\lambda)(\lambda-E_{2k}).
\eeq
($\mu$ is pure imaginary inside gaps and we choose $\tilde\mu$ to be real,
therefore we have the constant $i$ in (\ref{S2-tmu-def})).

Since the residues $r_k$ are pure imaginary, it is convenient to
introduce the real quantities $\tilde r_k(x,t)$ as follows:
\beq
\label{S2-tr-def}
r_k(x,t)=2i\tilde r_k(x,t).
\eeq
From (\ref{S2-r_k}) it follows:
\beq
\label{S2-tr_k}
\tilde r_k=\tilde\mu_k\prod\limits_{j\ne k}
\frac{\sqrt{\left|(\lambda_k-E_{2j})(\lambda_k-E_{2j+1})\right|}}
{\left|(\lambda_k-\lambda_j)\right|}.
\eeq

\subsection{Isoperiodic deformations of the spectral curve}

Finite-gap solutions corresponding to generic surfaces are quasiperiodic
in $x$ and $t$. To construct $x$-periodic solutions with a prescribed period,
it is necessary to put nontrivial additional constraints on the Riemann 
surface. These constraints for the KdV equation were formulated in 
\cite{MO} in terms of conformal maps. After minor modifications, the 
approach of \cite{MO} can be applied to the defocussing NLS equation but, 
for our purposes, an alternative
approach, suggested in \cite{GS}, is more convenient. The main idea of
\cite{GS} is to use the so-called {\bf isoperiodic deformations}, i.e.
ordinary differential equations on the moduli space of Riemann surfaces such
that all $x$ quasiperiods of the potential are conservation laws. Therefore,
if we have one periodic potential, we can construct new solutions
integrating these deformations. The starting point can be chosen in the
neighborhood of the zero potential using the perturbation theory.

Isoperiodic deformations arose in \cite{Kr1}, \cite{EFMS}, \cite{Sch96}
in the form of ODE's with a rather complicated transcendental right-hand side.
In \cite{GS} it was shown that, by extending the phase space,  these flows
 can be transformed to a very simple form with a rational right-hand side.

Consider the following collection of $g+1$ flows of the branch points $E_j$ 
and of the projections $\alpha_j$  of the quasimomentum differential zeroes
$E_{2j}<\alpha_j < E_{2j+1}$:

\beq
\label{S2-flow1}
\frac{\partial E _{j}}{\partial \tau_{k}}
=\frac{c_{k}}{E _{j}-\alpha _{k}} \mbox{ with }
j=0,\ldots ,2g+1, \ \ \ k=0,\ldots,g;
\eeq
\beq
\label{S2-flow2}
\frac{\partial \alpha _{j}}{\partial \tau_{k}}
=\frac{c_{k}}{\alpha _{j}-\alpha _{k}} \mbox{ with }
j=0,\ldots ,k-1,k+1,\ldots g;
\eeq
\beq
\label{S2-flow3}
\frac{\partial \alpha _{k}}{\partial \tau_{k}}
=\frac{1}{2} \sum_{j=0}^{2g+1}
\frac{c_{k}}{E _{j}-\alpha _{k}}-\sum_{j\neq k}
\frac{c_{k}}{\alpha _{j}-\alpha _{k}},
\eeq
where
\beq
\label{S2-ck}
c_{k}=\sqrt{\left|
\frac{\prod\limits_{j=0}^{2g+1}(\alpha _{k}-E _{j})}{
\prod\limits_{j\neq i} (\alpha _{k}-\alpha _{j})^{2}}
\right| } .
\eeq
Here $\tau_k$ denotes the parameter of the flow associated with the
point $\alpha_k$.

In \cite{GS} the following properties of equations
(\ref{S2-flow1})-(\ref{S2-flow3}) were proved:
\begin{lemma}
\begin{enumerate}
\item These flows preserve all $x$-quasiperiods of the corresponding NLS 
equation solutions. In particular, if the starting solution is $x$-periodic 
with the period $2L$, it remains $2L$-periodic in $x$.
\item These flows commute pairwise.
\item Assume that the parameters $\tau_k$ are normalized by the following 
condition: they tend to 0 if the lengths of all gaps tend to 0. Then
\beq
\label{S2-deftau}
p(\alpha_k^{\pm})=\mp i\tau_k, \ \ \mbox{mod}(\pi/L).
\eeq
\item The variation of the quasimomentum by these flows is given by
\beq
\label{S2-vardp}
\frac{\partial p(\gamma,\vec\tau)}{\partial \tau_k} =
-\frac{c_k}{\lambda-\alpha_k}
\frac{\partial p(\gamma,\vec\tau)}{\partial \lambda}.
\eeq
\end{enumerate}
\end{lemma}

\section{Asymptotics of the spectrum for discontinuous periodic profiles}

In the Fourier theory for periodic smooth functions it is
well-known that the coefficients $\hat f_k$ of the series decay faster than 
any power of $k$. If the function $f(x)$ is piecewise smooth and
discontinuous at a finite number of points $x_1$, \ldots, $x_n$ ,
then the coefficients decay as $1/k$ and the leading terms are:
\beq
\label{S3-Fourier}
\hat f_k= \frac{1}{2\pi i k} \sum\limits_{j=1}^n \delta_j
e^{-\frac{2\pi i}{2L}k x_j} + O\left(\frac{1}{k^2}\right).
\eeq
Here $2L$ is the period of the function and $\delta_j$ is the jump
at the point $x_j$: $\delta_j=f(x^+_j)-f(x^-_j)$, where $f(x^{\pm})=
\lim~f(x\pm |\epsilon |),~\epsilon\to 0$.
In addition, if the right and left derivatives
of $f(x)$ coincide at all discontinuity points, then the corrections
are of order $1/k^3$.

The periodic spectral transform can be treated as a nonlinear analog
of the discrete Fourier transform. In the small potential limit
the lengths of the gaps are $|2\hat f_k|$ and the phase of
$\hat f_k$ is encoded in the position of the divisor.

Since the Dirichlet IBV problem is reformulated as an odd
$2L$-periodic Cauchy problem with jumps $2v_0(t)$ at the points
$2nL$ and jumps $-2v_L(t)$ at the points $(2n+1)L$, then we
have to find a spectral characterization of this kind of
discontinuous profiles.

\begin{proposition}
\label{S3-prop1}
Let the potential $q(x)$ be odd, $2L$-periodic and  sufficiently
smooth outside the discontinuity points $nL$ and have the
jumps $2v_0(t)$, $-2v_L(t)$ at the points $2nL$, $(2n+1)L$
respectively. Then the odd character of $q(x)$ implies the following symmetry
of the spectrum:
\beq
\label{S3-symm}
E_k=-E_{-k},~~~\lambda_k=-\lambda_{-k},~~~k\in {\mathbb Z}
\eeq
and the above discontinuities imply the following asymptotics:
\beq
\label{S3-asympt}
\begin{array}{c}
E_{2k}=\frac{\pi k}{2L} + \frac{I_1}{\pi k} - \frac{|d_k|}{\pi k}+
O\left(\frac{1}{k^3}\right),\\
E_{2k+1}=\frac{\pi k}{2L} + \frac{I_1}{\pi k} + \frac{|d_k|}{\pi k}+
O\left(\frac{1}{k^3}\right),\\
\lambda_k=\frac{\pi k}{2L} + \frac{I_1}{\pi k} +
\frac{\Im d_k}{\pi k}+O\left(\frac{1}{k^3}\right),\\
\alpha_k=\frac{\pi k}{2L} + \frac{I_1}{\pi k} +
O\left(\frac{1}{k^3}\right),
\end{array}
\eeq
where
\beq
\label{S3-asympt2}
\begin{array}{c}
d_k=d_k(t)=v_0(t)-(-1)^{k}v_L(t), \\
I_1(t)=\int\limits_0^Ldx|u(x,t)|^2
\end{array}
\eeq
and $v_0(t)=u(0,t),~v_L(t)=u(L,t)$.
\end{proposition}
The absence of $O(1/k^2)$ corrections, due to the symmetry 
(\ref{S3-symm}), will be essential for us.

The idea of the proof is the following: in each continuity interval
equation (\ref{S2-lax1x}) can be asymptotically diagonalized by the
gauge transformation
\beq
\Psi(\lambda,x)=G(\lambda,x)\Psi_1(\lambda,x),
\eeq
where $G(\lambda,x)$ is an asymptotic series in $1/\lambda$. All
coefficients of $G(\lambda,x)$ can be computed explicitly in terms
of $q(x)$ and its derivatives. Taking into account the discontinuities
at $0$, $L$ and $2L$, one gets the following representation for
the transition matrix in the interval $[0,2L]$:
\beq
T(\lambda)=G(\lambda,0^-)e^{iL p_2(\lambda)\sigma_3 } G^{-1}(\lambda,L^+)
G(\lambda,L^-)e^{iLp_1(\lambda)\sigma_3} G^{-1}(\lambda,0^+),
\eeq
where $p_{1,2}(\lambda)$ are the averages on the intervals $[0,L]$
and $[L,2L]$ respectively of the first component of the diagonalized
$U$ matrix.

By Lemma~\ref{S2-l-spectrum} the branch points are determined by the
equation
$$
\tr T(\lambda)=\pm 2
$$
and the divisor at $x=0$ is determined by the equation
$$
T_{11}(\lambda)+T_{21}(\lambda)-T_{12}(\lambda)-T_{22}(\lambda)=0.
$$
The asymptotics (\ref{S3-asympt}) can be obtained expanding these
equations in inverse powers of $\lambda$.

\begin{lemma}
\label{S3-l-rs}
Let $\Gamma$ be an infinite genus spectral curve with
real branch points $E_k$ such, that
\beq
\label{S3-asympt3}
E_{2k}=\frac{\pi k}{2L} +O\left(\frac{1}{k}\right), \ \ \
E_{2k+1}=\frac{\pi k}{2L} +O\left(\frac{1}{k}\right)
\eeq
and each oval $a_k$ contains exactly one divisor point $\gamma_k$.
Then the quantities $r_k$ and  $\tilde r_k$ given by (\ref{S2-r_k}) 
and (\ref{S2-tr_k}) are well-defined.
\end{lemma}
The proof is rather straightforward.

\begin{center}
\mbox{\epsfxsize=9cm \epsffile{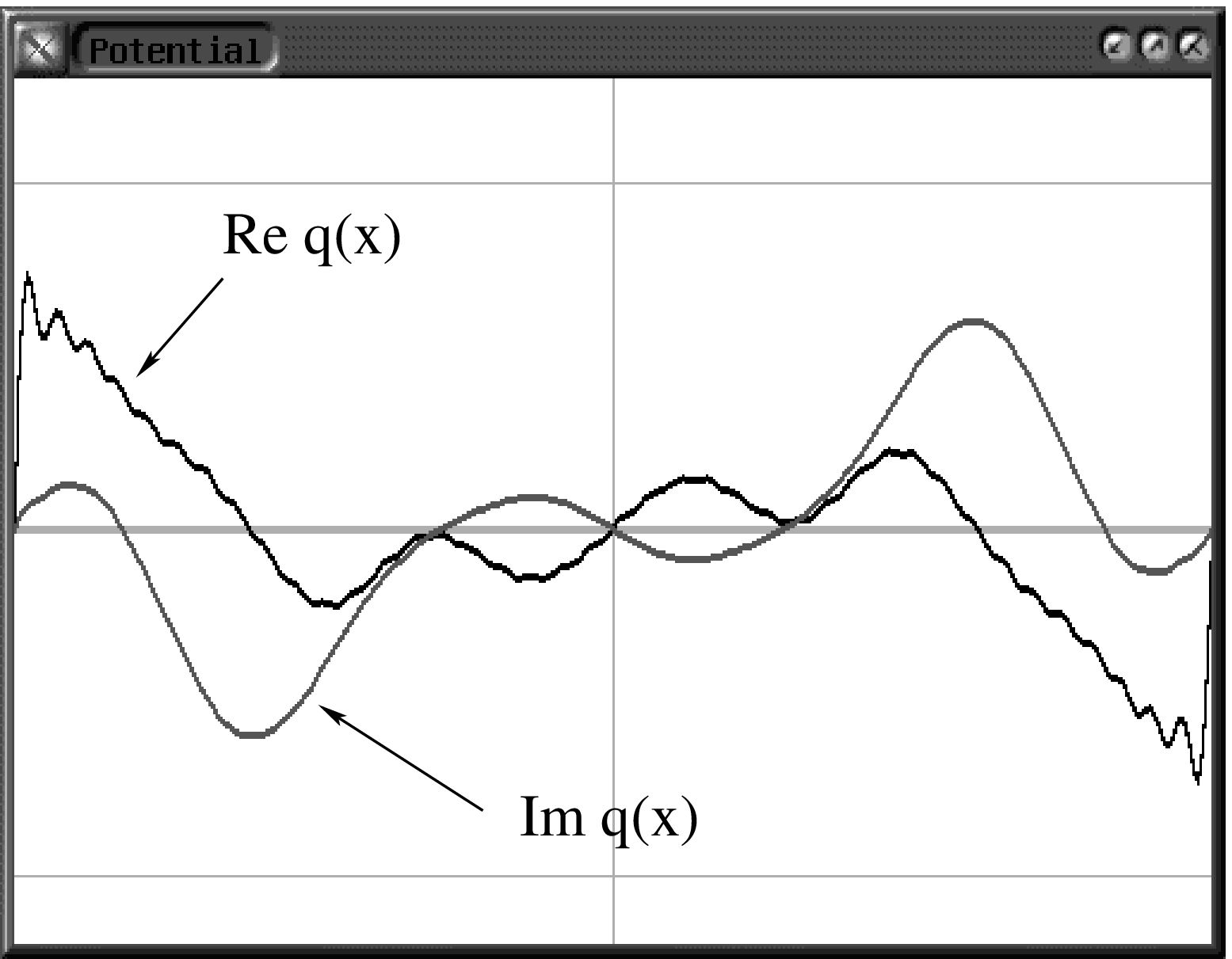}}

Fig 2.

This potential has 80 gaps (it corresponds to 40 Fourier harmonics).
\end{center}

\section{Time evolution of the spectral data}

In this section we derive the equations describing the time evolution
of the spectral data -- the spectral curve and the divisor -- for the
Dirichlet problem with zero at one end

\beq
\label{S4-IBV1}
u(0,t)=v_0(t), \ \ u(L,t)=v_L(t)=0, \ \ u(x,0)=u_0(x),\ \ \ 0\le x\le L.
\eeq
Let
\beq
\label{S4-BC1}
v_0(t)=A(t)e^{iB(t)},
\eeq
where $A(t)$ and $B(t)$ are real functions.

The calculations show that the evolution of the divisor is described by
the sum of two separately divergent vector fields. But if the boundary
function $v_0(t)$ is real, both divergences simultaneously vanish. This 
suggests to use the following gauge:
\beq
\label{S4-gauge}
q(x,t)=\tilde q(x,t) e^{iB(t)}
\eeq
and to calculate the Dubrovin equations for the function $\tilde q(x,t)$.
It satisfies the following equation:
\beq
\label{S4-NLSg}
i\tilde q_t(x,t) = \tilde q_{xx}(x,t)  - 2  |\tilde q(x,t)|^2 \tilde q(x,t)
+\dot B(t) \tilde q(x,t)
-2 A(t) \sum\limits_{n\in{\mathbb Z}}\delta'(x-2 Ln).
\eeq

The right-hand side of (\ref{S4-NLSg}) is a sum of 3 vector fields:
\beq
\frac{\partial \tilde q}{\partial t}= D_{s} \tilde q  +
 D_{g} \tilde q + D_{f} \tilde q,
\eeq
where
\beq
\label{S4-sflow}
D_{s} \tilde q=
-i\tilde q(x,t)_{xx}  + 2 i |\tilde q(x,t)|^2 \tilde q(x,t),
\eeq
\beq
\label{S4-gflow}
D_{g} \tilde q=
-i \dot B(t)  \tilde q(x,t),
\eeq
\beq
\label{S4-fflow}
D_{f} \tilde q=
2 i A(t) \sum\limits_{n\in{\mathbb Z}}\delta'(x-2Ln).
\eeq

Starting from now we denote by $\lambda_k(x,t)$ the divisor 
corresponding to the gauged potential $\tilde q(x,t)$.
It is natural to write the evolution of the branch points, of the zeroes 
of the quasimomentum differential and of the divisor as the sum of these 3 flows:

\beq
\label{S4-fflow-E}
\frac{d E_k(t)}{dt}= D_{s} E_k(t)+
D_{g} E_k(t)+D_{f} E_k(t),
\eeq
\beq
\label{S4-fflow-alpha}
\frac{d \alpha_k(t)}{dt}= D_{s} \alpha_k(t)+
D_{g} \alpha_k(t)+D_{f} \alpha_k(t),
\eeq
\beq
\label{S4-fflow-l}
\frac{\partial \lambda_k(0,t)}{\partial t}=
D_{s}\lambda_k(0,t)+D_{g}\lambda_k(0,t)+D_{f}\lambda_k(0,t).
\eeq

The action of the vector fields $D_s$ and
$D_g$ can be obtained as the $g\rightarrow\infty$
limit of the finite-gap formulas, but it is necessary to check
the convergence of these expression. From the results mentioned 
in Section 2 and, in particular, from equations (\ref{S2-Dubrt}),
(\ref {S2-Dubrph}), we have
\beq
\label{S4-sflow-E}
D_s E_k(t)=D_g E_k(t)=D_s \alpha_k(t)=D_g \alpha_k(t)=0,
\eeq
\beq
\label{S4-sflow-l}
D_s \lambda_k(0,t)=
4\left[\frac12\sum\limits_{j}E_j^2(t) -
\sum\limits_{j\ne k} \lambda_j^2(0,t)
\right] \tilde r_k(0,t),
\eeq
\beq
\label{S4-gflow-l}
D_g \lambda_k(0,t) =
-\frac{d B(t)}{d t} \tilde r_k(0,t).
\eeq
\begin{lemma}
Let the spectral data have the asymptotics (\ref{S3-asympt}) of
Proposition~\ref{S3-prop1}, with $d_k(t)=A(t)\in{\mathbb R}$. Then
the right-hand side of (\ref{S4-sflow-l}) is well-defined.
\end{lemma}
The proof is straightforward.

The rest of the section is devoted to the proof of the following 
Main Theorem, describing the effect of the forcing on the dynamics 
of the spectral data.

\begin{theorem}
\label{S4-theorem1}
The effect of the forcing on the spectral data is described by the 
following vector fields:

\beq
\label{S4-E-flow}
D_f E_j(t) =
\sum\limits_k
\frac{\partial E_{j}}{\partial \tau_{k}} D_f \tau_k(t)=
\sum\limits_k\frac{c_{k}(t)}{E_{j}(t)-\alpha_{k}(t)} D_f \tau_k(t),
\eeq
\beq
\label{S4-alpha-flow}
D_f \alpha_{j}(t)=
\sum\limits_k
\frac{\partial \alpha_{j}}{\partial \tau_{k}} D_f \tau_k(t)=
\sum\limits_{k\ne j} \frac{c_{k}(t)}{\alpha_{j}(t)-\alpha _{k}(t)} 
D_f \tau_k(t)+
\eeq
$$
+\left[ \frac{1}{2} \sum_{k}
\frac{c_{j}(t)}{E_{k}(t)-\alpha_{j}(t)}-\sum_{k\neq j}
\frac{c_{j}(t)}{\alpha_{k}(t)-\alpha_{j}(t)}    
\right] D_f \tau_j(t),
$$
\beq
\label{S4-lambda-flow}
D_f\lambda_k(0,t)=
\left[4A(t)\lambda_k(0,t)\frac{M_k(t)+2}
{M_k(t)}+2A(t)^2\right]\tilde r_k(0,t), 
\eeq
where
\beq
\label{S4-mutl}
M_k=e^{4iLp(\gamma_k(0,t),t)}-1,
\eeq
\beq
\label{S4-DG2}
D_f \tau_k(t)=
\frac{-2\alpha_k(t)}{L}A(t)
\frac{1- \tilde\Xi(\alpha^+_k,0,t)\tilde\Xi(\alpha^-_k,0,t)}
{\tilde\Xi(\alpha^+_k,0,t)-\tilde\Xi(\alpha^-_k,0,t)},
\eeq
\beq
\label{S4-defXi}
\tilde\Xi(\gamma,x,t)=\frac{\tilde\mu}{ \lambda-\lambda_k(x,t)}
\prod\limits_{j\ne k}\frac
{\sqrt{(\lambda-E_{2j}(t)) (\lambda-E_{2j+1}(t)) } }
{|\lambda-\lambda_j(x,t)|}+
\sum\limits_{k}\frac{\tilde r_k(x,t)}{(\lambda-\lambda_k(x,t))}
\eeq
and $c_k$ are defined by (\ref{S2-ck}).

These vector fields can be transformed into an algebraic form 
by introducing the following extended phase space:
$\{E_k,\alpha_k,\lambda_k,\mu_k,M_k\}$. The equations describing the 
time evolution of the remaining parameters have the following form:

\beq
\label{S4-mu-flow}
\frac{d}{dt}\tilde\mu_k(0,t)=\frac{\tilde\mu_k(0,t)}{2}\left[
\frac{\frac{d}{dt}\lambda_k(0,t)-\frac{d}{dt} E_{2k}(t)}{ \lambda_k(0,t) - E_{2k}(t) }+
\frac{\frac{d}{dt}\lambda_k(0,t)-\frac{d}{dt} E_{2k+1}(t)}{ \lambda_k(0,t) - E_{2k+1}(t)}
\right],
\eeq
\beq
\label{S4-M-flow}
\frac{d}{dt} M_k(t)=4L(M_k(t)+1) 
\prod\limits_{j\ne k}\frac{|(\lambda_k(0,t)-\alpha_j(t))|}
{\sqrt{(\lambda_k(0,t)-E_{2j}(t)) (\lambda_k(0,t)-E_{2j+1}(t)) } }
\times
\eeq
$$
\times
\frac{\lambda_k(0,t)-\alpha_k(t)}{\tilde\mu_k(0,t)}
\left[\frac{d}{dt} \lambda_k(0,t) - \sum\limits_j \frac{c_j(t)}
{\lambda_k(0,t)-\alpha_j(t)}
D_f \tau_j(t) \right].
$$
\end{theorem}

{\bf Remark.} The above dynamical system contains infinite sums and products. 
Their convergence follows from direct calculations using the 
asymptotic formulas (\ref{S3-asympt}).

The proof of Theorem~\ref{S4-theorem1} is based on the following formula:
\beq
\label{S4-monvar}
\delta\hat T(\lambda,t) =\int\limits_0^{2L} \hat T(\lambda,2L,\tilde x,t)
\delta U(\lambda,\tilde x,t) \hat T(\lambda,\tilde x,0,t)d\tilde x.
\eeq
Let us introduce the following notations:
\beq
\label{S4-tildeT}
\tilde T(\gamma,t,t_0)=\tilde\Psi^{-1}(\gamma,0,t_0)
\hat T(\lambda,t) \tilde\Psi(\gamma,0,t_0),
\eeq
where $\tilde\Psi(\gamma,x,t)$ is the following
$2\times 2$ matrix solution of (\ref{S2-lax1x}) combined from
the Bloch solutions:
\beq
\label{S4-tildePsi}
\tilde\Psi(\gamma,x,t)=\left[ \Psi(\gamma,x,t) \Psi(\sigma\gamma,x,t)
\right]= \left(\begin{array}{cc}
\psi^1(\gamma,x,t) & \psi^1(\sigma\gamma,x,t) \\
\psi^2(\gamma,x,t) & \psi^2(\sigma\gamma,x,t)
\end{array}\right).
\eeq
By definition,
\beq
\tilde T(\lambda,t_0,t_0)=\left(\begin{array}{cc} e^{2iLp(\gamma,t_0)} & 0 \\
0 & e^{-2iLp(\gamma,t_0)}
\end{array}\right).
\eeq
A direct calculation shows that
\beq
\left. \delta\tilde T(\lambda,t,t_0)\right|_{t=t_0}=
\tilde T(\lambda,t_0,t_0) \int\limits_0^{2L}
\tilde\Psi^{-1}(\lambda,\tilde x,t_0)
\delta U(\lambda,\tilde x,t_0)
\tilde\Psi(\lambda,\tilde x,t_0)
d\tilde x.
\eeq
For the variation of the potential $\tilde q$ defined by
(\ref{S4-fflow}) we obtain
$$
\left. \delta\tilde T(\gamma,t,t_0)\right|_{t=t_0}=
\tilde T(\gamma,t_0,t_0) \frac1W \int\limits_0^{2L}
\left(\begin{array}{cc}
\psi^2(\sigma\gamma,\tilde x,t_0) & -\psi^1(\sigma\gamma,\tilde x,t_0) \\
-\psi^2(\gamma,\tilde x,t_0) & \psi^1(\gamma,\tilde x,t_0)
\end{array}\right)\times
$$
\beq
\label{S4-DtildeT}
\times \left(\begin{array}{cc}
i\delta\lambda  &
-2 A(t_0)[\delta'(\tilde x )+ \delta'(\tilde x -2L)] \delta_f \\
-2 A(t_0)[\delta'(\tilde x )+ \delta'(\tilde x -2L)] \delta_f
& -i\delta\lambda
\end{array}\right)\times
\eeq
$$
\times\left(\begin{array}{cc}
\psi^1(\gamma,\tilde x,t_0) & \psi^1(\sigma\gamma,\tilde x,t_0) \\
\psi^2(\gamma,\tilde x,t_0) & \psi^2(\sigma\gamma,\tilde x,t_0)
\end{array}\right)=
$$
$$
=\frac{1}{W}\left(\begin{array}{cc} e^{2iLp(\gamma,t_0)} & 0 \\
0 & e^{-2iLp(\gamma,t_0)}\end{array}\right)
\left(\begin{array}{cc} \Delta_{11} &  \Delta_{12} \\
 \Delta_{21} &  \Delta_{22} \end{array}\right)(\gamma,t_0),
$$
where $W$ is the Wronskian determinant
\beq
\label{S4-W}
W=\psi^1(\gamma,x,t_0)\psi^2(\sigma\gamma,x,t_0)-
\psi^2(\gamma,x,t_0)\psi^1(\sigma\gamma,x,t_0),
\eeq
\beq
\label{S4-Delta11}
\Delta_{11}=i\delta\lambda\int\limits_0^{2L}
\psi^1(\gamma,x,t_0)\psi^2(\sigma\gamma,x,t_0)+
\psi^2(\gamma,x,t_0)\psi^1(\sigma\gamma,x,t_0)dx+
\eeq
$$
+2A(t_0)\delta_f\int\limits_0^{2L}\left(
\psi^1(\gamma,x,t_0)\psi^1(\sigma\gamma,x,t_0)-
\psi^2(\gamma,x,t_0)\psi^2(\sigma\gamma,x,t_0)\right)
[\delta'(x)+\delta'(x-2L)]dx,
$$
\beq
\label{S4-Delta21}
\Delta_{21}=-2i\delta\lambda\int\limits_0^{2L}
\psi^1(\gamma,x,t_0)\psi^2(\gamma,x,t_0)dx+
\eeq
$$
+2A(t_0)\delta_f\int\limits_0^{2L}\left(
(\psi^2(\gamma,x,t_0))^2-(\psi^1(\gamma,x,t_0))^2 \right)
[\delta'(x)+\delta'(x-2L)]dx
$$
(we do not use the formulas for $\Delta_{12}$, $\Delta_{22}$
in our calculations).

Let us first calculate the deformation of the spectral curve. From
(\ref{S2-alphas}), (\ref{S2-deftau}) it follows that
$$
\left.\frac{d \tau_k}{d t}\right|_{t=t_0}=
\left.i\frac{d p(\alpha^+_k(t)),t)}{d t}\right|_{t=t_0} =
i\left.\frac{\partial p(\gamma,t)}{\partial t}\right|_
{{{\cal P}\gamma=\alpha_k(t_0)}\atop{t=t_0}}.
$$
Then, from (\ref{S4-DtildeT}), it follows that
$$
\left.\frac{d \tau_k}{d t}\right|_{t=t_0}=
\left.\frac{\Delta_{11}}{2LW}\right|_{\gamma=\alpha^+_k(t_0)}
$$
with $\delta\lambda=0$.

The quadratic combination of eigenfunctions in (\ref{S4-Delta11})
is periodic, therefore we have
\beq
\label{S4-DG1}
\Delta_{11}=2A(t_0)\times
\eeq
$$
\times\int\limits_{-\epsilon}^{\epsilon}\left(
\psi^1(\alpha^+_k,x,t_0)\psi^1(\alpha^-_k,x,t_0)-
\psi^2(\alpha^+_k,x,t_0)\psi^2(\alpha^-_k,x,t_0)\right)
\delta'(x)dx=
$$
$$
=-2A(t_0)\left.\left[
\psi^1(\alpha^+_k,x,t_0)\psi^1(\alpha^-_k,x,t_0)-
\psi^2(\alpha^+_k,x,t_0)\psi^2(\alpha^-_k,x,t_0)\right]'\right|_{x=0}=
$$
$$
=-4i\alpha_k(t_0)A(t_0)\left[
\psi^1(\alpha^+_k,0,t_0)\psi^1(\alpha^-_k,0,t_0)+
\psi^2(\alpha^+_k,0,t_0)\psi^2(\alpha^-_k,0,t_0)\right].
$$

Using the fact that the Wronskian does not depend on $x$, the
definition (\ref{S2-defpsi}) and the arbitrariness of $t_0$,
we obtain equation (\ref{S4-DG2}).
Using the isoperiodic deformation formulas 
(\ref{S2-flow1})-(\ref{S2-flow2}), we finally obtain the evolution equations
(\ref{S4-E-flow}), (\ref{S4-alpha-flow}) of the spectral curve.

Let us calculate now the evolution of the divisor with respect to the
vector field $D_f$. Combining  Lemma~\ref{S2-l-spectrum} with the
definition of $\tilde T$ we obtain the following property of the
divisor points $\lambda_k$: the vector
$\left(\begin{array}{c} 1 \\ 0 \end{array}\right)$
remains the eigenvector of $\tilde T$ after variation,
or equivalently, $\Delta_{21}$=0. Therefore
\beq
\label{S4-DL1}
D_f\lambda_k(0,t)=iA(t)\frac{\cal N}{\cal D},
\eeq
where
$$
{\cal N}=-
\int\limits_0^{2L}\left(
(\psi^2(\gamma_k(0,t),x,t))^2-(\psi^1(\gamma_k(0,t),x,t))^2 \right)
[\delta'(x)+\delta'(x-2L)]dx=
$$
$$
=\int\limits_0^{2L}\left(
(\psi^2(\gamma_k(0,t),x,t))^2-(\psi^1(\gamma_k(0,t),x,t))^2 \right)'
[\delta(x)+\delta(x-2L)]dx,
$$
$$
{\cal D}=\int\limits_0^{2L}
\psi^1(\gamma_k(0,t),x,t)\psi^2(\gamma_k(0,t),x,t)dx.
$$
We used here the definiton of the divisor points $\gamma_k$ by the
condition $\psi^1(\gamma_k(x,t),x,t)+\psi^2(\gamma_k(x,t),x,t)=0$.
It is convenient to normalize the function $\Psi(\gamma,x,t)$
near the point $\gamma_k(0,t)$ by assuming
\beq
\label{S4-norm}
\psi^1(\gamma,0,t)-\psi^2(\gamma,0,t)=2.
\eeq
As a corollary we obtain
\beq
\label{S4-norm2}
\psi^1(\gamma_k(0,t),0,t)=1, \ \ \psi^2(\gamma_k(0,t),0,t)=-1,
\eeq
\beq
\label{S4-norm3}
\psi^1(\gamma,0,t)+\psi^2(\gamma,0,t)=\frac{2}{\Xi(\gamma,0,t)}.
\eeq

Let us calculate ${\cal N}$:
$$
\left((\psi^2(\gamma,x,t))^2-(\psi^1(\gamma,x,t))^2 \right)'=
$$
$$
=-2i\lambda(\psi^2(\gamma,x,t))^2+(\psi^1(\gamma,x,t))^2)-2i
(\tilde q(x,t)+\overline{\tilde q(x,t)})\psi^1(\gamma,x,t)\psi^2(\gamma,x,t);
$$
therefore
\beq
\label{S4-N1}
{\cal N}=-2i\lambda_k(0,t)\int\limits_0^{2L}\left(
(\psi^2(\gamma_k(0,t),x,t))^2+(\psi^1(\gamma_k(0,t),x,t))^2
\right)[\delta(x)+\delta(x-2L)]dx-
\eeq
$$
-2i\int\limits_0^{2L}
(\tilde q(x,t)+\overline {\tilde q(x,t)})\psi^1(\gamma_k(0,t),x,t)\psi^2(\gamma_k(0,t),x,t)
[\delta(x)+\delta(x-2L)]dx.
$$

If $f(x)$ is a continuous function, then
\beq
\label{S4-reg1}
\int\limits_0^{\epsilon} f(x)\delta(x)dx=\frac{1}{2}f(0),
\eeq
where $\epsilon$ is any positive constant. Therefore the first
integral in (\ref{S4-N1}) is well-defined:
\beq
\label{S4-num1}
\int\limits_0^{2L}\left(
(\psi^2(\gamma_k(0,t),x,t))^2+(\psi^1(\gamma_k(0,t),x,t))^2
\right)[\delta(x)+\delta(x-2L)]dx=
\eeq
$$
=e^{4iLp(\gamma_k(0,t))}+1.
$$
But the second integral depends on the regularization of the
$\delta$-function. In our problem the $\delta$-term arises as the
compensation of the discontinuity in $q(x,t)$. Therefore it is
natural to write, near the point $x=0$,
$$
\delta(x)=\frac{1}{2A(t)} \Re\tilde q_x(x,t) + \mbox{regular part}.
$$
We use also the fact that the function $\Im \tilde q(x,t)$ 
is continuous and vanishes at the point $x=0$; therefore
$$
\int\limits_0^{2L}
2 (\Re \tilde q(x,t) )\psi^1(\gamma_k(0,t),x,t)\psi^2(\gamma_k(0,t),x,t)
[\delta(x)+\delta(x-2L)]=
$$
\beq
\label{S4-num2}
=\lim\limits_{\epsilon\rightarrow0+}\left[\int\limits_0^{\epsilon}+
\int\limits_{2L-\epsilon}^{2L}\right]\frac{1}{A(t)}\Re\tilde q(x,t)
\Re\tilde q_x(x,t)
\psi^1(\gamma_k(0,t),x,t)\psi^2(\gamma_k(0,t),x,t)dx=
\eeq
$$
=\frac{1}{2}[e^{4iLp(\gamma_k(0,t))}-1]A(t).
$$
Combining (\ref{S4-num1}) and (\ref{S4-num2}) we obtain:
\beq
\label{S4-num3}
{\cal N}=-2i\lambda_k(0,t)\left[e^{4iLp(\gamma_k(0,t))}+1\right]-
iA(t)[e^{4iLp(\gamma_k(0,t))}-1].
\eeq

Let us now calculate ${\cal D}$:
$$
{\cal D}=\lim\limits_{\gamma'\rightarrow\gamma_k(0,t)}
\frac{1}{2}\int\limits_0^{2L}
\psi^1(\gamma_k(0,t),x,t)\psi^2(\gamma',x,t)+
\psi^1(\gamma',x,t)\psi^2(\gamma_k(0,t),x,t)dx=
$$
$$
=\frac{i}{2}\lim\limits_{\gamma'\rightarrow\gamma_k(0,t)}
\left.\left[\frac{
\psi^1(\gamma_k(0,t),x,t)\psi^2(\gamma',x,t)-
\psi^1(\gamma',x,t)\psi^2(\gamma_k(0,t),x,t)}
{\lambda'-\lambda_k(0,t)}
\right]\right|_0^{2L}=
$$
$$
=\frac{i}{2}\left[e^{4iLp(\gamma_k(0,t))}-1\right]\times
$$
$$
\times
\left.\frac{\partial}{\partial\lambda'}\left[
\psi^1(\gamma_k(0,t),0,t)\psi^2(\gamma',0,t)-
\psi^1(\gamma',0,t)\psi^2(\gamma_k(0,t),0,t)
\right]\right|_{\gamma'=\gamma_k(0,t)}=
$$
$$
=\frac{i}{2}\left[e^{4iLp(\gamma_k(0,t))}-1\right]
\left.\frac{\partial}{\partial\lambda'}\left[
\psi^1(\gamma',0,t)+\psi^2(\gamma',0,t)
\right]\right|_{\gamma'=\gamma_k(0,t)}.
$$
Taking into account (\ref{S4-norm3}) we obtain
\beq
\label{S4-den1}
{\cal D}=\left[e^{4iLp(\gamma_k(0,t))}-1\right]
\frac{i}{r_k(0,t)}
\eeq
and, finally, equation (\ref{S4-lambda-flow}).

Equation (\ref{S4-mu-flow}) can be obtained by direct differentiation of
(\ref{S2-tmu-rel}) and formula (\ref{S4-M-flow}) follows immediately 
from (\ref{S2-vardp}) and (\ref{S4-mutl}).

\section{Concluding remarks}
\begin{enumerate}
\item In our method we study the IBV problem (\ref{DfNLS}),
(\ref{IBV1}) through the following steps:
\begin{enumerate}
\item From the initial condition $u_0(x)$ we build its odd periodic 
extension $q(x,0)$ as in (\ref{IBV2}) and then we construct,
through the direct periodic spectral transform summarized in 
Section~\ref{Section2.1}, the initial spectral data 
$\{E_k(0),\alpha_k(0),\lambda_k(0,0),$ \break $\mu_k(0,0), 
M_k(0)\}$.
\item Solving the system of ODEs (\ref{S4-fflow-E}) - 
(\ref{S4-lambda-flow}),
(\ref{S4-mu-flow}), (\ref{S4-M-flow}) with the above initial
conditions, we obtain the evoluted spectral data 
$\{E_k(t),\alpha_k(t),\lambda_k(0,t),\mu_k(0,t),M_k(t)\}$.
\item\label{stepc}
Solving the $x$-Dubrovin equations (\ref{S2-Dubrx})
we obtain, from $\lambda_k(0,t)$,  $\lambda_k(x,t)$.
\item \label{stepd}
Using  formula (\ref{S2-potent}) we reconstruct the 
potential $q(x,t)$, whose restriction to the segment
$[0,L]$ gives the solution $u(x,t)$ of the IBV problem (\ref{DfNLS}),
(\ref{IBV1}).
\item It is well-known that the $x$-Dubrovin equations can be linearized,
at fixed $t$, on the Jacobi variety constructed from the spectral data 
$\{E_k(t),\gamma_k(0,t)\}$. Correspondingly, the solution $q(x,t)$ 
can be written explicitly in terms of Riemann $\theta$-functions. 
This is an alternative procedure to the above steps (\ref{stepc}),
(\ref{stepd}). However we remark that this Jacobi variety depends
on $t$ in a highly non-trivial way, in contrast to the integrable
cases.
\end{enumerate}
\item Althouh the $x$-Dubrovin equations can be linearized on the
Jacobi variety and the solutions can be written down explictly in terms
of Riemann $\theta$-functions, it turns out that, from the numerical
point of view, it is often (and, in particular, in our case) more
convenient to integrate the Dubrovin equations directly. This idea
was communicated to one of the authors (P.G.) by A.Osborne long times 
ago, and it was used by the author in his numerical experiments. 
\item We believe that our approach be the proper nonlinear analog 
of the discrete sine-Fourier transform method, which allows one to 
solve the Dirichlet problem on the segment for the linear 
Schr\"odinger equation.
\item We consider this approach - in which the original
IBV problem is reformulated as a forced periodic IBV problem that,
without forcing, is known to be integrable - as a convenient way
to treat the original IBV problem. One of the reasons is the
following. It is well-known that, to analyse a perturbed (by a
forcing) nonlinear equation, one has to calculate the spectral
decomposition of the forcing in terms of the eigenfunctions of
the linearized equation. But, for  soliton equations, this
decomposition can be explicitly written in terms of the auxiliary
spectral problem. Therefore the calculation of the time evolution
of the spectral data due to the forcing can be considered as the
proper nonlinear analog of the Fourier analysis.
\item A detailed study of the system of ODEs 
(\ref{S4-fflow-E}) - (\ref{S4-lambda-flow}),
(\ref{S4-mu-flow}), (\ref{S4-M-flow}) will be the subject of
future investigation. Preliminary numerical experiments seem to
show that, starting with zero initial condition and applying 
a smooth localized forcing, only a finite number of 
nonlinear modes remain essentially non-zero when the 
external forcing vanishes. Correspondingly, only a finite subset
of the above system of ODEs is essential for gaining a 
correct picture of the process.
\item The method presented in this paper can obviously be
applied to study the general Dirichlet, Neumann and mixed 
problems for the focussing and defocussing NLS equations on the
segment. These general results will be presented in subsequent
works.
\end{enumerate}

\bigskip

\noindent
{\large\bf Acknowledgments} 
We acknowledge useful discussions with A.Degasperis and
S.V.Manakov.

\end{document}